\documentclass[reprint,twocolumn,longbibliography]{revtex4}
\usepackage[english]{babel}
\usepackage[utf8]{inputenc}
\usepackage{amsmath}
\usepackage{amssymb}
\usepackage{amsthm}
\usepackage{mathtools}
\usepackage{dsfont}
\usepackage{graphicx}
\usepackage{hyperref}
\usepackage[draft,inline,nomargin]{fixme} \fxsetup{theme=color}

\usepackage{xcolor}

\newcommand{\ket}[1]{\left|{#1}\right\rangle}
\newcommand{\bra}[1]{\left\langle{#1}\right|}
\newcommand{\braket}[2]{\langle{#1}|{#2}\rangle}
\newcommand{\ketbrad}[1]{\left|{#1}\rangle\!\langle{#1}\right|}

\newcommand{\zme}{Z^{\mathrm{ME}}~}

\DeclareMathOperator{\Tr}{Tr}

\DeclarePairedDelimiter\ceil{\lceil}{\rceil}
\DeclarePairedDelimiter\floor{\lfloor}{\rfloor}

\begin{document}
\title{Certified answers for ordered quantum discrimination problems}
\date{\today}
\author{Esteban Mart\'inez Vargas}
\email{Esteban.Martinez@uab.cat}
\affiliation{F\'isica Te\`orica: Informaci\'o i Fen\`omens Qu\`antics, Departament de F\'isica, Universitat Aut\`onoma de Barcelona, 08193 Bellatera (Barcelona) Spain}
\author{Ramon Mu\~noz-Tapia}
\email{Ramon.Munoz@uab.cat}
\affiliation{F\'isica Te\`orica: Informaci\'o i Fen\`omens Qu\`antics, Departament de F\'isica, Universitat Aut\`onoma de Barcelona, 08193 Bellatera (Barcelona) Spain}
\begin{abstract}
We investigate the quantum state discrimination task for sets of linear independent  
pure states with an intrinsic ordering. This structured discrimination 
problems allow for a novel scheme that provides a certified level of error, that is, 
answers that never deviate from the true value more than a specified distance and hence a control  of
the desired quality of the results.
We obtain an efficient semidefinite program and also find a general lower bound valid for any error distance that only requires the knowledge of optimal minimum error scheme.
We apply our results to the quantum change point and quantum state anomaly detection cases.
\end{abstract}
\maketitle
\section{Introduction}
State discrimination plays a fundamental role in quantum information sciences as it 
determines the capacity of quantum systems to carry information. The task consists in 
identifying in which of some known set of states a system was prepared by some source. 
If the possible states are mutually orthogonal this task can be done perfectly. However,
if the states are not mutually orthogonal the problem is  very nontrivial and it 
requires optimization with respect to some reasonable criteria. 
 
The most studied  discrimination schemes are minimum error (ME) and unambiguous discrimination (UD).
In ME after a measurement is performed on the system the experimenter must give an answer
about its state. Naturally, some of the answers will be erroneous, and  the optimal ME  strategy
is the one that yields the minimum probability of committing an error~\cite{Helstrom1976}. In contrast, in 
UD,  no errors are allowed, i.e, the answers of the experimenter must be absolutely certain. This can only be achieved  at the expense of permitting
inconclusive measurement outcomes. The optimal strategy is the one that minimizes the probability of inconclusive answers. It is known that 
UD is only possible for sets of linearly independent states \cite{chefles1998unambiguous}. For mixed states UD is also possible as long as they 
 do not have identical supports~\cite{rudolph2003unambiguous}.

Some extensions of these fundamental schemes have also been considered.  Discrimination 
with maximum confidence~\cite{croke2006maximum}  can be applied to states that are not necessarily independent 
and  can be regarded as a generalized UD strategy. Strategies that interpolate between ME and UD 
have also been studied \cite{bagan2012Qrate}. In those a given maximum value for the error probability 
(or equivalently a maximum value for inconclusive probability)  is enforced. Varying 
this value yields a continuous set of strategies between UD 
(or maximum confidence) and ME.

Despite being such a fundamental task, analytical solutions for optimal discrimination 
schemes in the multi-hypothesis case remains  a challenge (see \cite{singal2019MEindependent} for 
recent developments). Essentially only the two state \cite{Helstrom1976}  and symmetric 
states cases \cite{ban1997optimum,barnett2001symmetric,krovi2015symmetric} have been solved (see  \cite{Barnett09QD,Chefles2000QDReview,Bae_2015}  for 
reviews on state discrimination).

In this work we consider a novel multi-hypothesis scheme for sources that prepare 
states with intrinsic structure. In particular, we consider  linear independent states that can be 
represented as a linear chain (see Fig~\ref{fig:schemefig}) of $n$ local states. This type of sources includes  
the interesting cases of change point \cite{Sentis2016QCP,QCPUnAmb,Sentis2019Online} and state anomaly detection 
 \cite{Identification2018Skotiniotis} problems. 
In these structured sources the hypotheses are labelled by some position in the chain,
Hence the errors  have a natural 
distance, i.e., we can have have a one-site error, two-site error, etc-.,  if the outcome 
of the protocol is an answer that is at distance of one, two, etc., units  from the site labelling 
the true hypothesis. This scheme is interesting not only from the theoretical point of 
view, but also for practical purposes. In many circumstances not any error can be tolerated, 
however small deviations from the true hypothesis may have only a  limited impact on 
our decisions.  So, it  may prove useful to find optimal schemes under the constraint 
that no outcome can differ from the true hypothesis more than a given threshold distance 
$\Delta$. Doing so, we have certified answers that will not spoil decisions that we may 
take upon the outcome of the protocol. We therefore call this scheme certified answer discrimination (CAD).
Also if
    we relax the UA condition and  
allow some errors, the success 
probability of guessing the correct hypothesis  can  increase substantially as we will show.  For 
$\Delta=0$ we recover the UD scheme while  for $\Delta=n-1$ we get the ME scheme, 
thus CAD also provides an interpolation between UD and ME. 
The  interpolating scheme discussed in~\cite{bagan2012Qrate} also yield a significant increase in the success probability, 
but, contrasting  the CAD scheme, it  may give  erroneous answers that are very far from the
true value. As it will become clear, CAD is a more natural scheme, closer to the notion of
Hamming distances between states (i.e, the sum of positional mismatches \cite{Nielsen2011QCQ}).
 
In this paper we give a convenient and efficient semidefinite program (SDP) \cite{VandenbergheSDP1996,Eldar2002ASDP,
watrous_2018}
formulation of CAD schemes for linearly independent states. The SDP also  enables us to 
find an analytical  lower bound for the probability of success for any allowed error distance
$\Delta$.
 Interestingly, this lower bound only requires to calculate the  ME 
success probability, It provides an approximation on how much the 
success probability is  reduced as we increase the requirements on the quality of the answers of the discrimination protocol.

The paper is organized as follows. In the next Section we present the CAD scheme and 
its SDP formulation. In Section III we obtain a lower bound for the success probability 
for any value of $\Delta$. In Section IV we apply our results to 
the paradigmatic case of the change point and also discuss the state anomaly detection problem.  
Section IV contains the conclusions of our findings. We also include an appendix with some
technical details.
\section{Certified answer discrimination $\Delta$-schemes}

\begin{figure}[h]
    \includegraphics[width=0.48\textwidth]{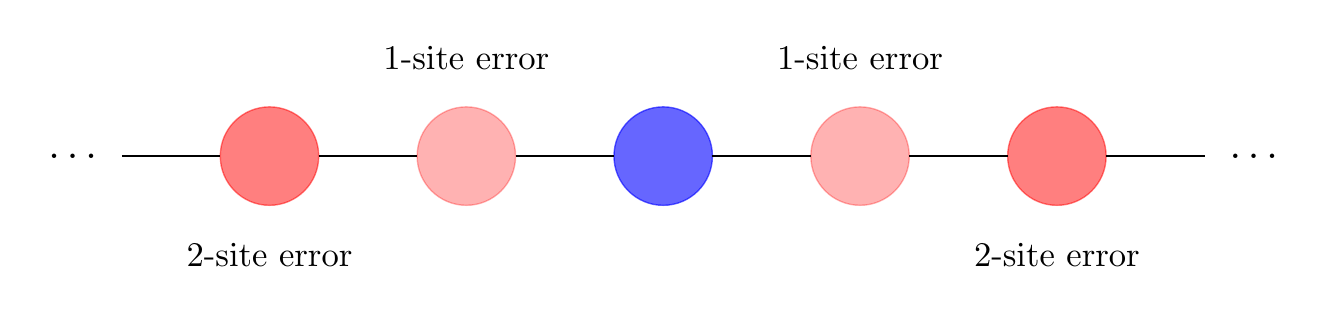}
    \caption{Structure of the source states. The position of the blue dot labels the state.}
    \label{fig:schemefig}
\end{figure}

Consider a quantum state multi-hypothesis discrimination problem where the source 
quantum states have an intrinsic ordering such as a one dimensional chain as depicted 
in Fig.~\ref{fig:schemefig}. In this case it is possible to define a natural distance 
between the states. 

If we are given a state $\ket{\Psi_k}$, where $k$  is the position that  labels the 
state, our aim is to find a measurement, generally given by a Positive Operator Value  
(POVM), that returns the value $k$ with the highest probability. The POVM has to 
satisfy the additional constraints that no errors beyond some distance $\Delta$ can be 
committed. For states given by a chain of  $n$ states, the POVM contains $n+1$ 
elements $\{E_k\geq 0\}_{k=0}^n$, where $E_0=\openone -\sum_{k=1}^n E_k$ is the 
element corresponding to an  inconclusive answer. As in UD this element has to 
be introduced in order to satisfy the constraints. Naturally as $\Delta$ 
increases, i.e, more and more  type of errors are allowed, we have  
$\bra{\Psi_k} E_0\ket{\Psi_k}\to 0$ for  $k=1,2,\ldots,n$. 

The optimization problem can be written as the following SDP:
    \begin{equation}
    \begin{aligned}
        & \underset{E}{\text{maximize}}
        & & \frac{1}{n}\sum_{i=1}^n\bra{\Psi_i}E_i\ket{\Psi_i} \\
        & \text{subject to}
        & & \bra{\Psi_j}E_i\ket{\Psi_j}=0~~\forall|i-j|> \Delta\\
        & & & \sum_{i=1}^n E_i\leq \mathds{1} \\
        & & & E_i\geq 0~\forall i,
    \end{aligned}
    \label{eq:uglysdp}
\end{equation}
where for simplicity  we assume that the prior probability is the same for 
all source states. We will also assume that the source states are linear 
independent, as naturally happens in the examples
considered here (see section IV). Observe that  
each value $\Delta=0,1,2,\ldots,n-1$  defines  a discrimination scheme that we will 
call a $\Delta$-scheme.
Note also that $E_0$ is a slack variable that it is  taken into account by the 
inequality $\sum_{i=1}^n E_i\leq \openone$ in the POVM condition.

For a given value of  $\Delta$ we have 
a probability of success $P_s^\Delta$, a probability of error $P_e^\Delta$ and a probability of inconclusive outcome $P_I^\Delta$, and they satisfy
the unitarity condition $P_s^\Delta+P_e^\Delta+P_I^\Delta=1$. 
The value $\Delta=0$ corresponds to the unambiguous case for which the error probability vanishes, $P_e^{\Delta=0}=0$, and  the outcome can either perfectly identify the state or 
be inconclusive, but not erroneous. For
$\Delta=n-1$ the are no constraints on the errors and we recover the minimum error 
scheme, i.e  the inconclusive probability  vanishes, $P^{\Delta=n-1}_I=0$.  As we will 
see later, the minimum error limit can be effectively achieved for much smaller values 
of $\Delta$.

\begin{figure}
    \centering
    \includegraphics[width=0.4\textwidth]{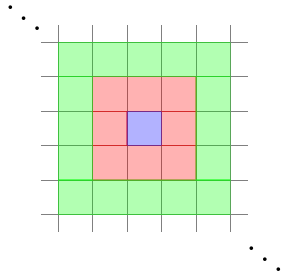}   
    \caption{Example of the structure of $F_r^\Delta=R^\dagger E_r^\Delta R$. The matrices $F_r^\Delta$  have dimensions
        $n\times n$ and their non-vanishing elements are depicted as colored boxes.
     For $\Delta=0$ (no errors), the  central blue box is the only non-vanishing element.  For $\Delta=1$ (one error)
      the non-vanishing elements are contained in the red $3\times3$ block, and in the
    $5\times5$ green block  for $\Delta=2$, etc. The remaining entries of $F_r^\Delta$  are all zero.}
    \label{fig:povmelementR}
\end{figure}

If the source states are linearly independent, we can transform the 
SDP (\ref{eq:uglysdp}) into an equivalent and  more useful program. 
From the  $n$ linearly independent estates $\{\ket{\Psi_i}\}_{i=1}^n$ we construct 
the $R$ matrix,
\begin{equation}
    R = \sum_{i=1}^n|\Psi_i\rangle\langle i |,
\end{equation}
where $\ket{i}$ is any orthonormal basis (note that linear independence implies that $R$ is 
invertible) 
and consider the new operators 
$F_r^\Delta=R^\dagger E_r^\Delta R$. Observe that the diagonal elements of $F_r^\Delta$ are the expectation values 
$\bra{\Psi_i} E_r^\Delta|\ket{\Psi_i}=[F_r^\Delta]_{i,i}$. Thus, the first constraint in Eq.~(\ref{eq:uglysdp})
translates into the condition that  all diagonal elements $[F_r^\Delta]_{i,i}$ vanish except those with $|i-r|\leq\Delta$.  Note also that
 $E_r^\Delta\geq 0 \rightarrow F_r^\Delta\geq 0$~\cite{bhatia1996matrix}.  The off-diagonal terms $[F_r^\Delta]_{i,j}$ are then also constrained by  
 positivity, and hence 
 we  have  $[F_r^\Delta]_{i,j}=0$ for $|r-i|> \Delta$ and
$|r-j|> \Delta$. The structure of the matrix $F_r^\Delta$ is illustrated in Fig.~(\ref{fig:povmelementR}).

\begin{figure}
    \centering
    \includegraphics[width=0.4\textwidth]{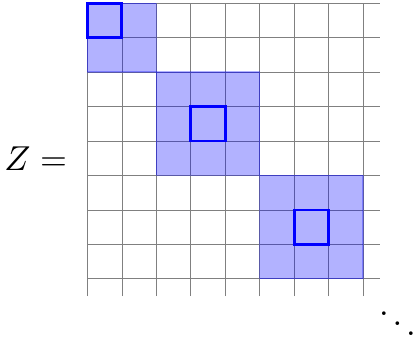}
    \caption{Structure of the matrix variable $Z$ for $\Delta=1$.
    The blue boxes correspond to the free matrix elements and the blank ones are fixed to be zero. The highlighted boxes are the elements
        that appear in the objective function $(1/n)\Tr [ZA]$ of Eq.~\eqref{eq:nicesdp}.}
    \label{fig:cajasvar}
\end{figure}

The second constraint in Eq.~(\ref{eq:uglysdp}) can be recast as 
\begin{equation}
\label{eq:constraint}
    G - \sum_{r=1}^nF_r^\Delta\geq 0,
\end{equation}
by applying  the matrix $R^{\dagger}$  on the left and the matrix $R$ on the right.
Here $G=R^\dagger R$ is the Gram matrix~\cite{HornMA2012} whose elements are
\begin{equation}
    G_{i,j}=\braket{\Psi_i}{\Psi_j}.
\end{equation}
Thus the SDP~(\ref{eq:uglysdp}) is transformed onto
\begin{equation}
    \begin{aligned}
        & \underset{Z}{\text{maximize}}
        & & \frac{1}{n}\Tr[ZA] \\
        & \text{subject to}
        & & \Phi_\Delta[Z] \leq G \\
        & & & Z\geq 0.
    \end{aligned}
    \label{eq:nicesdp}
\end{equation}
The matrix variable $Z$ has a block diagonal structure containing the non-vanishing 
elements of $F^\Delta_r$.  In Fig.~\ref{fig:cajasvar}  we explicitly depict it for 
$\Delta=1$. The elements highlighted are the ones that appear in the objective 
function $\Tr[ZA]$. The constant matrix $A$ depends on the number $n$ of hypothesis 
and maximum distance  $\Delta$ of the allowed errors (we do not add these labels to 
avoid cluttering too much the notation). Matrix $A$ "selects'' the elements  of the 
matrix variable $Z$ that have to be optimized, i.e., the central elements of the $Z$ 
blocks. For $\Delta=1$ one has  $A = \text{diag}\{1,0,0,1,0,0,1,\ldots,1\}$  and 
$Z$ and $A$ are $(3n-2)\times(3n-2)$ matrices.
The generalization for any $\Delta$ is straightforward. Note that the appearance 
    of the Gram matrix $G$ in the second constraint of \eqref{eq:nicesdp}
    showcases that all the discrimination properties of sets of linearly independent 
    states are encapsulated in the Gram matrix.

\begin{figure}
    \centering
    \includegraphics[width=0.48\textwidth]{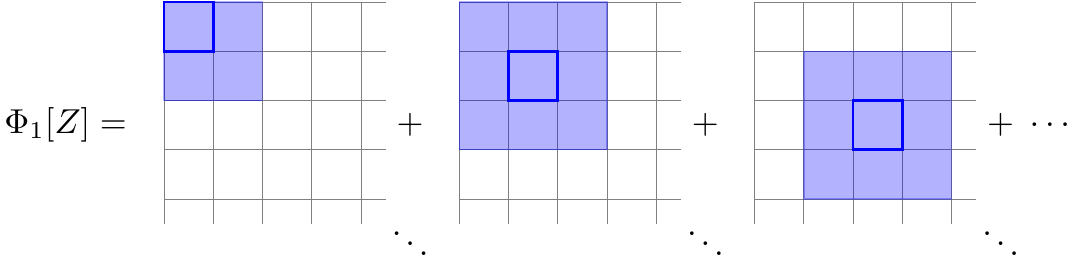}
    \caption{The correspondent map takes the non-zero parts of the variable $Z$ and
        accommodates it in $n\times n$ matrices with zeros in the remaining places. Observe
        that matrix sum is defined only for matrices of the same dimensions.}
    \label{fig:mapeo}
\end{figure}

The linear map  $\Phi_{\Delta}$ that incorporates the constraints \eqref{eq:constraint}  can be regarded as the action of two linear maps:
$ \Phi_\Delta = \Phi^2\circ\Phi^1_\Delta$.
The first map, $\Phi^1_\Delta$, embeds each block into a $n\times n$ sub-matrix and 
pads the remaining elements with zeros. The embedding is such that the $k$'th sparse 
sub-matrix has  the  central (highlighted) element in the $k$th position of the 
diagonal, as  can be seen in Fig~\ref{fig:mapeo}. With all the sub-matrices 
 we have an $n^2\times n^2$ block diagonal matrix.
The second map, $ \Phi^2$, adds the sub-matrices  
to get a final $n\times n$ matrix, also illustrated in 
Fig.~\ref{fig:mapeo}. Notice that this map is independent of $\Delta$.

We note that the variable $Z$ from SDP (\ref{eq:nicesdp})
has dimensions $[n(2\Delta+1)-\Delta(\Delta+3)]\times[n(2\Delta+1)-\Delta(\Delta+3)]$ 
which is significantly lower than $n^2\times n^2$ 
of the original SDP (\ref{eq:uglysdp}). The size of the variables is 
similar only for $\Delta\rightarrow n$. 
However, as we will see in the quantum change point, the ME limit can be effectively reached  for small values of $\Delta$, and then the number of variables remains low for all meaningful values of $\Delta$.

There is no general mathematical method for solving a given SDP analytically, only problems with high degree
of symmetry are known to be solvable. In some cases the primal or the dual version of the SDP can suggest 
an ansatz that may provide the solution (see \cite{QCPUnAmb} for a nice example). 
Therefore, any understanding of the form of the solutions of SDPs at hand is of interest. The transformation of the
SDP made above proves to be beneficial not only for 
the numerical advantage but also to obtain insight into how the probability of success behaves
in the intermediate regime between unambiguous and minimum error schemes.
In particular,  it enables us that find useful analytical lower bounds
of  the probability of success for any $\Delta$ that we discuss in the next section.

\section{A lower bound for $P_s^{\Delta}$}

\label{sec:lowerbound}
The main idea is to obtain a feasible solution of the SDP (\ref{eq:nicesdp}). Any 
ansatz matrix $\tilde{Z}$ that satisfies the constraints of an SDP is by construction 
a lower bound to the optimal solution. The method depends heavily on having previously 
solved the ME scheme, i.e. we have at our disposal the success probability  
$P_s^{\mathrm{ME}}$, and the corresponding  $\zme$. Fortunately, in many cases
the minimum error scheme can be computed or well approximated with a square 
root measurement~\cite{Hausladen94PrettyGood,Hausladen96Classical}.


As discussed in previous section the mapping $\Phi_\Delta[Z]$ in the SDP (\ref{eq:nicesdp}) can be understood as two step mapping that first transforms 
the variable $Z$ into
a $n^2\times n^2$ variable that has zeros in appropriate places and a second step that sums all the individual blocks into a $n\times n$ matrix.
 If we only apply the first map, we get the following SDP:
\begin{equation}
\begin{aligned}
        & \underset{Z}{\text{maximize}}
        & & \frac{1}{n}\Tr[ZA] \\
        & \text{subject to}
        & & \Phi_\Delta^1[Z] \leq \zme \\
        & & & Z\geq 0.
    \end{aligned}
    \label{eq:restrictsdp}
\end{equation}
Observe that any variable $Z$ that satisfies $\Phi_\Delta^1[Z] \leq \zme$ also satisfies  $\Phi_\Delta[Z] \leq G$ (just apply the 
map $\Phi^2$, on both sides of the first inequality). Hence, any feasible solution of the SDP (\ref{eq:restrictsdp}) is in the feasible set of the SDP (\ref{eq:nicesdp}),  but not vice versa, and it provides  a lower bound for the probability of success.

For simplicity, let us  call $\Phi_\Delta^1[Z]=Z^\Delta$ and $[Z_r^{\Delta}]_{i,j}$ the
$i,j$ element of its $r$-th sub-matrix $Z_r^{\Delta}$. The positivity condition $\zme - Z^\Delta \geq 0$ in Eq.~\eqref{eq:restrictsdp}
implies that any principal minor of $\zme - Z^\Delta $ has to be positive~\cite{HornMA2012}. 
\begin{figure}
    \centering
    \includegraphics[width=0.48\textwidth]{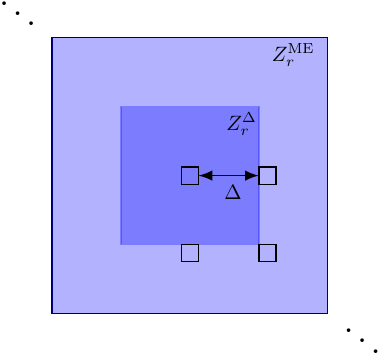}
    \caption{We depict a specific block  $r$ of  $\zme-Z^{\Delta}$. The light
        blue block corresponds to $Z^{\mathrm{ME}}_r$ and the darker one to
        $Z^{\Delta}_r$. The small black boxes show the elements of the minor 
    of interest to obtain the  bound  \eqref{eq:guessatom}. }
    \label{fig:boxproof}
\end{figure}
To get a bound in terms of the known $\zme$, the elements of the principal minor have to be outside the central blocks
of $Z^\Delta$, as depicted in  Fig.~\ref{fig:boxproof}.
The choice of this minor is such that it contains only one non-vanishing diagonal element of $Z^\Delta$ and three remaining elements are at a $\Delta$ distance and hence take the (known) $\zme$ values. We take the minimum distance $\Delta$ as larger distances will give less stringent bounds.

The positivity condition then gives
\begin{eqnarray}
    \left([\zme_i]_{i,i} - [Z^{\Delta}_{i}]_{i,i}\right)[\zme_{i}]_{i+\Delta+1,i+\Delta+1}\geq\nonumber\\
    \left|[\zme_{i}]_{i,i+\Delta+1}\right|^2. 
\end{eqnarray}
Using the fact that the arithmetic mean is bigger than the geometric mean we finally have that
\begin{align}
    \big([\zme _i]_{i,i}- & [Z^{\Delta}_{i}]_{i,i}\big)+ [\zme_i]_{i+\Delta+1,i+\Delta+1}\nonumber \\
     & \geq 2\left|[\zme_{i}]_{i,i+\Delta+1}\right|. 
    \label{eq:ineqminor}
\end{align}
As we will be dealing with problems having some symmetry it is convenient to choose 
this lower minor for the first $\ceil*{n/2}$ and the corresponding  upper minor for the 
rest of blocks. For these upper minors we get 
the same inequality 
\eqref{eq:ineqminor} with the change $\Delta\to -\Delta$.

In order to calculate the bound of the success probability only the diagonal elements 
of $\tilde{Z}$ have to be specified. The best choice is to take them to 
saturate  the inequalities \eqref{eq:ineqminor}, i.e.,
\begin{align}
    [\tilde{Z}_i]_{i,i} =\begin{cases}   [\zme_{i}]_{i,i}-H_i (\Delta) &  \mbox{for}\  1\leq i\leq \ceil*{n/2} \\
                                 [\zme_{i}]_{i,i}-H_i(-\Delta)  & \mbox{for}\  i>\ceil*{n/2} 
                                \end{cases} , 
        \label{eq:guessatom}
\end{align}
where 
\begin{equation}
   H_i (\Delta)=2\left |[\zme_{i}]_{i,i+\Delta+1}\right| - [\zme_i]_{i+\Delta+1,i+\Delta+1} .
        \label{eq:HDelta}
\end{equation}
Adding all the terms in Eq.~\eqref{eq:guessatom},  the lower bound $\tilde{P}_s $ for the success probability reads
\begin{align}
   P_s^{\Delta} \geq & \tilde{P}_s  =  P_s^{\mathrm{ME}}\nonumber \\
 -& \frac{1}{n}\left[\sum_{i=1}^{\ceil*{n/2}}H_i(\Delta) +\sum_{i=\ceil*{n/2}+1}^{n}H_{i}(-\Delta)\right],
    \label{eq:genPguess}
\end{align}
which depends only on $\zme$. The bound (\ref{eq:genPguess}) has two parts, the first is just the success probability 
of the minimum error case (i.e., the unrestricted case), while the second takes into account how much this 
value is diminished by the additional constraints imposed by the value $\Delta$. 
The main virtue of this bound is that given the solution for the minimum error case it provides an expression on how 
much this probability is lessened by  increasing the quality of the answers, i.e., by reducing the maximum allowed distance of the answers
to the true state.

\section{Applications}
In this section we apply our findings to two paradigmatic multi-hypothesis cases. We first discuss
the Quantum Change Point (QCP) problem~\cite{Sentis2016QCP,QCPUnAmb,Sentis2019Online} and then briefly  discuss  
 the Quantum State Anomaly Detection (QSAD) problem~\cite{Identification2018Skotiniotis}.

\begin{figure}[h]
    \centering
    \includegraphics[width=0.48\textwidth]{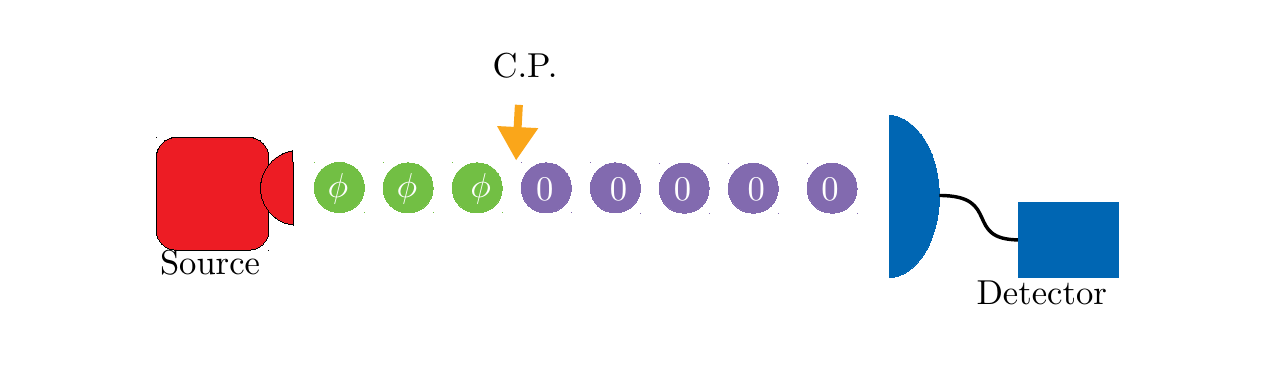}
    \caption{A machine produces a signal state and suddenly it produces another
        signal. Our task is to determine by measurements the exact moment when this
        change happens.}
    \label{fig:qcp}
\end{figure}

The QCP problem is depicted in figure (\ref{fig:qcp}).
A source prepares systems in a default state $|0\rangle$
for some time and suddenly  it changes and prepares systems in a 
mutated state 
$|\phi\rangle$. Both states are assumed to be known and the change is also assumed to 
occur at any time with the same probability.  The total number of systems is $n$. 
The goal is to identify the position of the mutation with the highest probability. 
This is a multi-hypothesis case  for which the optimal ME and UD probabilities of 
success are known~\cite{Sentis2016QCP,QCPUnAmb}.  

The global states can be written as
\begin{equation}
    |\Psi_k\rangle = |0\rangle^{\otimes k-1}|\phi\rangle^{\otimes n-k+1}.
\end{equation}
The Gram matrix 
has elements $G_{i,j}=\braket{\Psi_i}{\Psi_j}=c^{|i-j|}$,
where $c=\braket{0}{\phi}$ and w.l.o.g. can be taken to be in the interval  $0\leq c \leq 1$. Note that for $c\neq 0,1$ the off diagonal elements of the 
$G$ decay exponentially as they depart from the diagonal , which shows that in the QCP the  Hamming distance between states
 is directly related to the overlap between states. 

\begin{figure}
    \centering
    \includegraphics[width=0.48\textwidth]{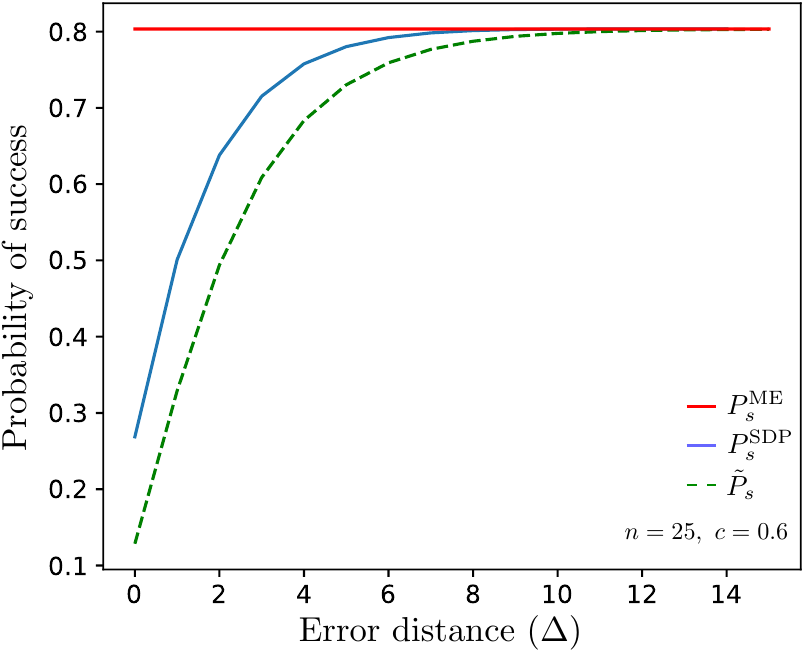}
    \caption{Probability of success versus the allowed error distance
            $\Delta$ for QCP  with $n=25$ and $c=0.6$. The blue solid curve are the exact numerical SDP values
            \eqref{eq:nicesdp}. The dotted green curve is the analytical lower bound \eqref{eq:PguessQs}. We also show as a reference the 
            red straight line with the value  of  the minimum
            error scheme.}
    \label{fig:Ps_ks}
\end{figure}

\begin{figure}[tbh]
    \centering
    \includegraphics[width=0.4\textwidth]{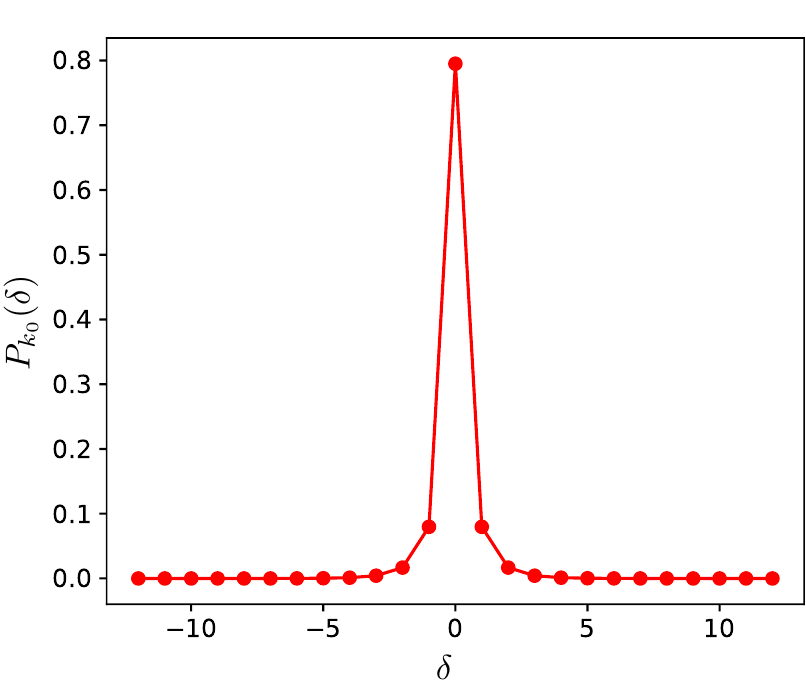}
    \caption{Outcome probability profile of the minimum error scheme of the 
            QCP. The parameter $\delta=\hat{k}-k_0$ is the distance of the output 
            guess $\hat{k}$ with respect to the position $k_0$ of the true  change 
            point. Here  $n=25$ , $c=0.6$ and we take the change point to occur at 
            the  central position $k_0=13$.}
    \label{fig:k-errors}
\end{figure}

The CAD scheme is particularly pertinent for this problem.  
It is reasonable to assume that here some deviations of the output guess from the true change point can be tolerated, but not too many  in order to avoid jeopardizing the validity of the identification task.  In Fig.~\ref{fig:Ps_ks} we show the success probability as a function of $\Delta$ as given by the SDP~\eqref{eq:nicesdp} for $c=0.6$ and $n=25$.
We note a remarkable  increase in the  success probability by just 
allowing one error deviation of the guess. The value of $P_s$  jumps almost a factor of two,  from 
$0.27$ for $\Delta=0$, to $0.50$ for $\Delta=1$. Also the inconclusive probability drops from 0.73 to 0.4, while only  10\% of the answers will be
erroneous (and just by one position). If these are counted as satisfactory answers, the total success probability goes up to 60\%.
We have checked that these values of the probabilities essentially remain constant for any $n>25$. 
We also observe that the probability of 
success stabilizes to the ME value for $\Delta\gtrsim 8$ (again this threshold value remains the same for larger values of $n$).
This just shows that the ME 
protocol effectively does not yield answers that are at distance greater than eight 
space units from the true state, as can explicitly  be seen in Fig.~\ref{fig:k-errors}. 

We next calculate the bound \eqref{eq:genPguess}. As discussed in previous section,  the bound requires to have the solution $\zme$, but, as shown in \cite{Sentis2016QCP}, for the QCP  can be very 
well approximated by he square root measurement, i.e.  by  a projective POVM 
$\{E_k=\ketbrad{m_k}\}_{k=1}^n$, with $S\ket{m_k}=\ket{\Psi_k}$ and $S=\sqrt{G}=\sum_k\sqrt{\lambda_k}\ketbrad{v_k}$ , where $\lambda_k$ and $\ket{v_k}$ are the eigenvalues and eigenvectors of $G$, respectively.

The matrix $\zme$ in terms of the square root $S$ simply reads 

\begin{equation}
\zme=\bigoplus_{k=1}^n \ketbrad{s_k} \ \ \mbox{with } \ \ \braket{m_l}{s_k}=S_{l,k},
\label{eq:zme-s}
\end{equation}

i.e. $\ket{s_k}$ are the column vectors of $S$.

The crucial point to obtain a useful bound is to prove that the elements of  $S$  away from the diagonal decay exponentially.
From the supplemental material of \cite{Sentis2016QCP} we have 
\begin{widetext}
\begin{equation}
    S_{k,l} \approx \frac{\sqrt{1-c^2}}{\pi}\int_0^\pi d\theta \frac{(\sin k\theta - c\sin(k-1)\theta)(\sin l\theta - c\sin(l-1)\theta)}{(1-2c\cos\theta+c^2)^{3/2}} .
    \label{eq:srmelapprox}
\end{equation}
\end{widetext}
After some straightforward algebra  Eq.~\eqref{eq:srmelapprox} reads
\begin{align}
    S_{k,l} &\approx \frac{\sqrt{1-c^2}}{4\pi}\int_{-\pi}^\pi d\theta \left[\frac{\cos(k-l)\theta-\cos(k+l)\theta}{(1-2c\cos\theta+c^2)^{1/2}}\right.\nonumber\\
        &+ \left. \chi(k+l,c)\right],
    \label{eq:srmelapproxfourier}
\end{align}
where $\chi(k+l,c)$ contains terms that oscillate rapidly and will be considered later (observe that the second term also oscillates more rapidly than the first).
We also note that the explicit terms terms shown in Eq.~\eqref{eq:srmelapproxfourier} correspond to the Fourier  series of
of the function 
\begin{equation}
 \mu(\theta,c) = \frac{1}{(1-2c\cos\theta+c^2)^{1/2}},
\end{equation}
so  we consider
\begin{equation}
    \widehat{\mu}(k,c) = \int_{-\pi}^{\pi}\mu(\theta,c)e^{ik\theta}d\theta
    \label{ec:fourierterms}
\end{equation}
for $k\in\mathds{N}$. We prove in  Appendix \ref{app:A}  that $\widehat{\mu}(k,c)$
exhibits an exponential decay in $k$ given by
\begin{equation}
        |\widehat{\mu}(r,c)|\leq M_0(c)e^{k\log(c)}.
        \label{eq:mu-bound}
    \end{equation}
where $M_0=  \int_{-\pi}^\pi\mu(\theta,c)d\theta$.
The other terms included in $\chi(k+l,c)$ of Eq.~\eqref{eq:srmelapproxfourier})are proportional to $ \mu^3(r,c)$ and can be tackled in a similar fashion.
Including the term proportional to $\cos(k+l)\theta$  and the terms coming from $\chi(k+l,c)$  we get 
\begin{align}
   S_{k.l} \leq \frac{\sqrt{1-c^2}}{4\pi}\left( M_0 e^{|k-l|\log (c)} 
 + \sum_{i=-1}^{2} M_ie^{(k+l+i)\log(c)}\right).
    \label{eq:srmelements}
\end{align}
We can now  calculate $\zme$ inserting \eqref{eq:srmelements} into  Eq.~\eqref{eq:zme-s}. We  further just take into consideration the (first) dominant term to obtain 
\begin{align}
    \left |[\zme_i]_{i,i\pm\Delta+1}\right| &\leq c\,e^{\Delta\log c}\left|[\zme_i]_{i,i}\right| \nonumber \\
    \left |[\zme_i]_{i\pm\Delta+1,i\pm\Delta+1}\right| &\leq c^2 \,e^{2\Delta\log c}\left|[\zme_i]_{i,i}\right| .
\end{align}
Finally from equation (\ref{eq:genPguess}) we get
\begin{equation}
    \tilde{P}_s \geq  (1- 2ce^{\Delta\log c}+ c^2e^{2\Delta\log c})P_s^{\mathrm{ME}},
    \label{eq:PguessQs}
\end{equation}
which shows that the success probability approaches  at least exponentially  $P_s^{\mathrm{ME}} $
for sufficiently large $\Delta$.  Note also that in the limit $c\to 0$ we recover the obvious result that $P^{\mathrm{ME}}=P^{\mathrm{UA}}$.
We show  the bound \eqref{eq:PguessQs} along with the exact numerical results in 
figure (\ref{fig:Ps_ks}). We observe that indeed the bound approaches the minimum 
error value for large $\Delta$. 
 
To end this section we study the Quantum State Anomaly Detection (QSAD) problem 
\cite{DallaPozza2015Optimality,Identification2018Skotiniotis} , which will provide 
some further insight of the features of the our certified answers protocol.  QSAD  
can be regarded as a simplified case of the QCP.  The source is assumed to 
    prepare systems in a given default sate $|0\rangle$, however one (and just one) 
    of the local systems was prepared in a different anomalous state $\ket{\phi}$. 
As in the QCP we assume both states to be known and  equal probability for the position 
of the anomalous state.  The task consists in identifying the position of the faulty 
state with the highest probability when a string of $n$ systems has been 
    prepared. Also here we may consider a 
    protocol that yields guesses not deviating more than $\Delta$ units from the true 
    position of the anomaly. 

The set of hypothesis is is given by
\begin{equation}
    |\Psi_k\rangle = |0\rangle^{\otimes k-1}|\phi\rangle|0\rangle^{\otimes n-k}.
\end{equation}
and again we define  $c = \braket{\phi}{0}$ that w.l.o.g. can be taken to be in the 
interval $0\leq c\leq1$.  Notice that we have a very simple Gram matrix in this case  
\begin{equation}
G_{i,j}=\braket{\psi_i}{\psi_j}=(1-c^2)\delta_{ij}+c^2.
\label{eq:qpeigram}
\end{equation} 
This Gram matrix  is circulant \cite{Gray2006Toeplitz} , and hence the  square root measurement is  optimal~\cite{DallaPozza2015Optimality,Sentis2016QCP}.
It is straightforward to find $S=\sqrt{G}$:
\begin{equation}
S_{i,j}=(a-b)\delta_{ij}+b
\label{eq:qsad}
 \end{equation}
 where 
 \begin{align}
 a=&\frac{\sqrt{1+(n-1)c^2}+(n-1)\sqrt{1-c^2}}{n}\nonumber \\
 b=&\frac{\sqrt{1+(n-1)c^2}-\sqrt{1-c^2}}{n}
 \end{align}
Note that the  success probability for the minimum error scheme is simply \cite{DallaPozza2015Optimality}
\begin{equation}
P_s^{\mathrm{ME}}=\frac{1}{n}\sum_{i=1}^n S_{i,i}^2=a^2.
\label{eq:minerr-qsad}
\end{equation}

\begin{figure}
      \centering
      \includegraphics[width=0.48\textwidth]{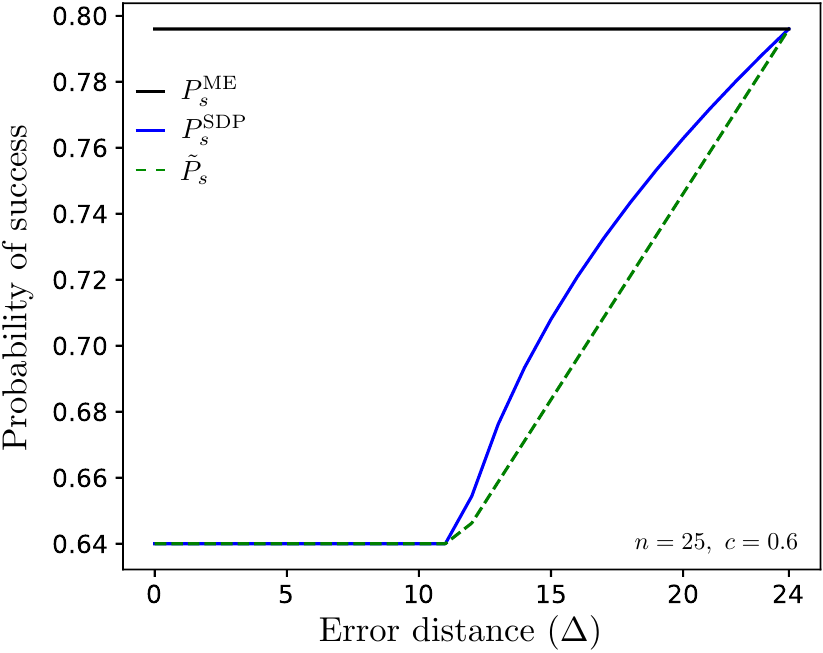}
      \caption{Probability of success against the
              error distance $\Delta$ for QSAD with $n=25$ and $c=0.6$. The blue solid line shows the 
          numerical results from SDP \eqref{eq:nicesdp} 
          and the dashed green line the lower bound Eqs. \eqref{eq:PUA-qsad}
          and \eqref{eq:pguessqpei2}. We also show as a reference a 
           black  straight line with  the value  of  the minimum
            error scheme.}
      \label{fig:derrorsqpei}
  \end{figure}

The fact that all source states have the same overlap, or equivalently have equal Hamming distance, makes the distance to the 
true anomaly a less natural parameter in this case and we  have different behaviors for $\Delta<\floor{n/2}$
and $\Delta\geq\floor{n/2}$. It is easy to convince oneself that the symmetry of 
the problem implies that the condition $\bra{\psi_i}E_j\ket{\psi_i}=0$ for 
$|i-j|\geq \Delta$ for any $\Delta<\floor{n/2}$ is in fact  equivalent to impose 
$\Delta=0$ . Whence for $0\leq \Delta<\floor{n/2}$ we have a constant probability of 
success,  as can be seen in Fig.~\ref{fig:derrorsqpei}, and the protocol is equivalent 
to unambiguous discrimination. It is interesting to calculate the bound 
\eqref{eq:genPguess} in this regime. We have 
\begin{align}
    [\zme_i]_{ii}  = a^2,   \ \ \ 
    [\zme_i]_{i,i\pm\Delta+1} & = ab \nonumber  \\
    [\zme_i]_{i\pm\Delta+1,  i\pm\Delta+1} = b^2 &. 
\end{align}
From equation \eqref{eq:genPguess}
we get 
\begin{equation}
   \tilde {P}_s= (a-b)^2=1-c^2.
   \label{eq:PUA-qsad}
\end{equation}
This value is exactly the  unambiguous success probability.  Notice that for $\Delta=0$,  the matrix $A$ in Eq.~\eqref{eq:nicesdp}  is $\openone_n$  and  that by symmetry $Z=z\openone_n$, with $z$ a real parameter. Then the  SDP reads
\begin{equation}
    \begin{aligned}
        & \text{maximize} 
        & & z \\
        & \text{subject to}
        & & z\openone_n \leq G \\
        & & & z\geq 0,
    \end{aligned}
    \label{eq:sdp-ua}
\end{equation}
which is the SDP for the minimum eigenvalue of $G$. From \eqref{eq:qpeigram} it is direct to obtain $z=1-c^2$, as expected.

For $\Delta\geq\floor{n/2}$ we can start having some  errors, and the success probability starts to increase from UA to ME as seen in 
Fig.~\ref{fig:derrorsqpei}.
We also see that the  lower bound~\eqref{eq:genPguess} in this regime  departs from the $P_s^{\mathrm{UA}}$ value. 
Now at least one block of $\zme$ can be completely covered  by $Z^\Delta$ which allows for larger contributions to the bound.
So  $[\tilde{Z}_j]_{ii}$ has some elements constrained to be $(a-b)^2$ and as $\Delta$ increases new ones equal to the larger value $a^2$.
Defining $d:=\Delta-\floor{n/2}$ and recalling that $P^{\mathrm{ME}}_s=a^2$ and $P_s^{\mathrm{UA}}=(a-b)^2$, we obtain from  Eq.~\eqref{eq:genPguess} 

    \begin{equation}
    \tilde{P}_s=
    \begin{cases*}
        \frac{n-(2d+1)}{n}P_s^{\mathrm{UA}}+\frac{2d+1}{n}P_s^{\mathrm{ME}} & for \text{$n$ odd} \\
        \frac{n-2d}{n}P_s^{\mathrm{UA}}+\frac{2 d}{n}P_s^{\mathrm{ME}} & for \text{$n$ even},
    \end{cases*}
    \label{eq:pguessqpei2}
\end{equation} 
which exhibits a nice linear behavior interpolating between UA and ME.

\section{Conclusions}
We have introduced a novel scheme of quantum discrimination for ordered hypothesis of 
linearly independent states that gives certified answers that do not depart from the 
true hypothesis more than a given distance $\Delta$. Our scheme may be of practical importance 
in cases where small deviations from the true hypothesis can be tolerated without 
compromising the effectiveness of the discrimination task. The scheme allows to tune at will  the quality versus the 
quantity of the answers.

We have shown that all the 
discrimination properties of a given set of hypotheses are contained in the Gram 
matrix of the set. We  have obtained a compact SDP  for  the optimal solution that 
can be solved very efficiently. We have also obtained a lower bound of the success 
probability for any value of the deviation  that only requires the knowledge of the 
minimum error solution. The bound gives an analytical expression of how much the 
minimum error success probability is reduced  as the maximum distance error $\Delta$ is  decreased.

We have applied our findings to the quantum change point problem and the quantum state anomaly detection. For the former, we have shown that allowing a small departure from the true change point  increases quite dramatically the success probability. We have computed the lower bound  
and shown that the increase of the success of probability is exponential in the allowed distance of the errors. 
%
 For the QSAD we see that 
up to $n/2$ the protocol is equivalent to unambiguous discrimination. 
The lower bound for $\Delta\geq n/2$ gives  a linear interpolation 
between UA and ME error protocols. 

Our scheme is versatile enough to address other interesting situations. For instance, 
one might consider non-symmetric errors, i.e the tolerated distance of forward and 
backward errors may be different. Also one can consider incompatibilities, i.e, given 
some hypothesis the protocol is required to avoid some specific answers. One 
important extension of our protocol would be to consider sets of linearly 
dependent and noisy states. The main difficulty here is how to extend the Gram matrix 
formalism in these settings.  We are currently exploring these scenarios.
 
\acknowledgments
We thank useful discussions with Gael Sent\'{\i}s.
We acknowledge the financial support of the Spanish MINECO, ref. FIS2016-80681-P 
(AEI/FEDER, UE), and Generalitat de Catalunya CIRIT, ref. 2017-SGR-1127. EMV thanks 
financial  support from CONACYT.
\appendix
\section{}
\label{app:A}
In this Appendix we prove  that 
the Fourier coefficients $\widehat{\mu}(k,c)$ of Eq.~\eqref{ec:fourierterms} decay exponentially
    with $k$ as
    \begin{equation}
        |\widehat{\mu}(k,c)|\leq M(c)e^{k\log(c)},
    \end{equation}
    where
    \begin{equation}
    M(c)= \int_{-\pi}^{\pi} \mu(\theta,c) d\theta.
    \end{equation}
\begin{figure}[htb]
        \centering
        \includegraphics[width=0.48\textwidth]{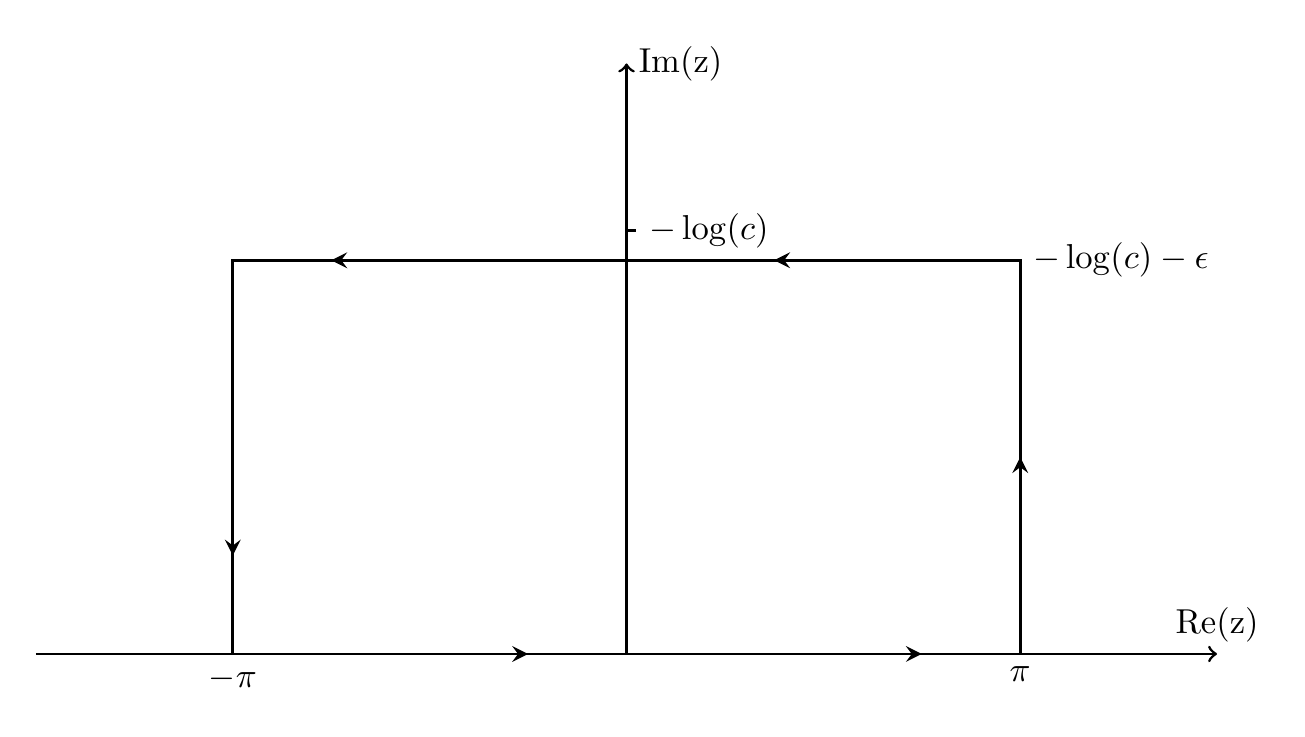}
        \caption{
        Path of the contour integral of $\mu(z,c)$  in the complex plane.}
        \label{fig:contour}
    \end{figure}
\begin{proof}
We first extend the function $\mu(\theta,c)$ to the complex plane as
\begin{equation}
    \mu(z,c) = \frac{1}{\sqrt{(c-e^{iz})(c-e^{-iz})}}.
    \label{ec:expeigval}
\end{equation}
If we take the principal branch of the logarithm  as a domain of $z\rightarrow\sqrt{z}$
the function $\mu(z,c)$, is analytic in $\mathds{C}/\{-i\log(c),i\log(c)\}$
because it is the composition of several analytic functions.
It is a known fact that the Fourier coefficients of analytic functions decay
exponentially~\cite{Katznelson2002}.
We next compute
\begin{equation}
    \widehat{\mu}(k,c)= \int_{-\pi}^{\pi}\mu(\theta,c)e^{ik\theta}d\theta,
    \label{ec:fourierterms-A}
\end{equation}
for $k\in\mathds{N}$. Notice that due to the symmetry $\mu(\theta,c)=\mu(-\theta,c)$, only the cosine term of $e^{ik\theta}$ survives.  We consider the contour integral in the complex plane shown in Fig.~\ref{fig:contour}.
We will call $\gamma^c$
 the part of the contour that does not lie in the real line.
        By analyticity of $\mu(z,c)$ in this region we have that,
    \begin{equation}
        \int_{-\pi}^\pi \mu(\theta,c)e^{ik\theta}d\theta + 
        \int_{\gamma^c} \mu(z,c)e^{ikz}dz = 0.
        \label{eq:contoureq}
    \end{equation}
   Notice that $\mu(\theta,c)\geq 0~\forall~\theta\in[-\pi, \pi]$ and $0\leq c\leq1$, i.e, 
  $\mu(\theta,c)=|\mu(\theta,c)|$. We also  
    see that  the contributions of the right and left  vertical sections of the path cancel out. Thus, we have 
    \begin{equation}
        \int_{-\pi}^\pi |\mu(\theta,c)|d\theta = 
        \int_{-\pi}^\pi |\mu(x+i(-\log(c)-\epsilon),c)|dx,
        \label{eq:truco1}
    \end{equation}
and  from Eq.~\eqref{eq:contoureq}  we get
    \begin{align*}
        \left|\int_{-\pi}^\pi \mu(\theta,c)e^{ik\theta}d\theta\right| &=  
        \left|\int_{\gamma^c} \mu(z,c)e^{ikz}dz \right|\\
        &\leq \int_{\gamma^c} \left|\mu(z,c)e^{ikz}\right|dz\\
        &=\int_{\gamma^c} \left|\mu(z,c)\right|e^{-ky}dz\\
        &= e^{k(\log(c)+\epsilon)}\\
        &\times \int_{-\pi}^{\pi}|\mu(x+i(-\log(c)-\epsilon),c)|dx\\
        &= M(\epsilon,c)e^{k(\log(c)+\epsilon)},
    \end{align*}
     where in going from the second to the third r.h.s expression we use the fact that the the right and left arms contributions of the contour $\gamma^c$
     cancel out. Note that the constant $M(\epsilon,c)$ does not depend on $k$. 
    Taking the limit $\epsilon\rightarrow0$ and recalling Eq.~\eqref{eq:truco1},  we get
    \begin{equation}
        |\widehat{\mu}(k,c)|\leq e^{k\log(c)}\int_{-\pi}^\pi\mu(\theta,c)d\theta,
    \end{equation}
   i.e., $ M(c)= \int_{-\pi}^{\pi} \mu(\theta,c) d\theta$. 
    \end{proof}

We can calculate in a completely analogous fashion the Fourier coefficients for 
other powers of $\mu(\theta,c)$. For instance, the function $\chi(k+l,c)$ in Eq.~\eqref{eq:srmelapproxfourier} 
includes terms proportional to $\mu^3(\theta,c)$ and these will also decay 
exponentially. 

All the elements of $S=\sqrt{G}$ of the QCP can thus be expressed as
\begin{align}
   S_{k,l} &\leq \frac{\sqrt{1-c^2}}{4\pi}\left( M_0 e^{|k-l|\log (c)}\right. \nonumber\\
    &+ \left.\sum_{i=-1}^{2} M_ie^{(k+l+i)\log(c)}\right),
    \label{eq:srmelements-A}
\end{align}
where $M_i$ are constants that only depend on $c$ and not on $k$ or $l$.

\bibliography{QuantumSD}

\begin{thebibliography}{27}%
\makeatletter
\providecommand \@ifxundefined [1]{%
 \@ifx{#1\undefined}
}%
\providecommand \@ifnum [1]{%
 \ifnum #1\expandafter \@firstoftwo
 \else \expandafter \@secondoftwo
 \fi
}%
\providecommand \@ifx [1]{%
 \ifx #1\expandafter \@firstoftwo
 \else \expandafter \@secondoftwo
 \fi
}%
\providecommand \natexlab [1]{#1}%
\providecommand \enquote  [1]{``#1''}%
\providecommand \bibnamefont  [1]{#1}%
\providecommand \bibfnamefont [1]{#1}%
\providecommand \citenamefont [1]{#1}%
\providecommand \href@noop [0]{\@secondoftwo}%
\providecommand \href [0]{\begingroup \@sanitize@url \@href}%
\providecommand \@href[1]{\@@startlink{#1}\@@href}%
\providecommand \@@href[1]{\endgroup#1\@@endlink}%
\providecommand \@sanitize@url [0]{\catcode `\\12\catcode `\$12\catcode
  `\&12\catcode `\#12\catcode `\^12\catcode `\_12\catcode `\%12\relax}%
\providecommand \@@startlink[1]{}%
\providecommand \@@endlink[0]{}%
\providecommand \url  [0]{\begingroup\@sanitize@url \@url }%
\providecommand \@url [1]{\endgroup\@href {#1}{\urlprefix }}%
\providecommand \urlprefix  [0]{URL }%
\providecommand \Eprint [0]{\href }%
\providecommand \doibase [0]{https://doi.org/}%
\providecommand \selectlanguage [0]{\@gobble}%
\providecommand \bibinfo  [0]{\@secondoftwo}%
\providecommand \bibfield  [0]{\@secondoftwo}%
\providecommand \translation [1]{[#1]}%
\providecommand \BibitemOpen [0]{}%
\providecommand \bibitemStop [0]{}%
\providecommand \bibitemNoStop [0]{.\EOS\space}%
\providecommand \EOS [0]{\spacefactor3000\relax}%
\providecommand \BibitemShut  [1]{\csname bibitem#1\endcsname}%
\let\auto@bib@innerbib\@empty
\bibitem [{\citenamefont {Helstrom}(1976)}]{Helstrom1976}%
  \BibitemOpen
  \bibfield  {author} {\bibinfo {author} {\bibfnamefont {C.~W.}\ \bibnamefont
  {Helstrom}},\ }\href@noop {} {\emph {\bibinfo {title} {Quantum detection and
  estimation theory}}}\ (\bibinfo  {publisher} {Academic press},\ \bibinfo
  {year} {1976})\BibitemShut {NoStop}%
\bibitem [{\citenamefont {Chefles}(1998)}]{chefles1998unambiguous}%
  \BibitemOpen
  \bibfield  {author} {\bibinfo {author} {\bibfnamefont {A.}~\bibnamefont
  {Chefles}},\ }\bibfield  {title} {\bibinfo {title} {Unambiguous
  discrimination between linearly independent quantum states},\ }\href@noop {}
  {\bibfield  {journal} {\bibinfo  {journal} {Physics Letters A}\ }\textbf
  {\bibinfo {volume} {239}},\ \bibinfo {pages} {339} (\bibinfo {year}
  {1998})}\BibitemShut {NoStop}%
\bibitem [{\citenamefont {Rudolph}\ \emph {et~al.}(2003)\citenamefont
  {Rudolph}, \citenamefont {Spekkens},\ and\ \citenamefont
  {Turner}}]{rudolph2003unambiguous}%
  \BibitemOpen
  \bibfield  {author} {\bibinfo {author} {\bibfnamefont {T.}~\bibnamefont
  {Rudolph}}, \bibinfo {author} {\bibfnamefont {R.~W.}\ \bibnamefont
  {Spekkens}},\ and\ \bibinfo {author} {\bibfnamefont {P.~S.}\ \bibnamefont
  {Turner}},\ }\bibfield  {title} {\bibinfo {title} {Unambiguous discrimination
  of mixed states},\ }\href@noop {} {\bibfield  {journal} {\bibinfo  {journal}
  {Physical Review A}\ }\textbf {\bibinfo {volume} {68}},\ \bibinfo {pages}
  {010301} (\bibinfo {year} {2003})}\BibitemShut {NoStop}%
\bibitem [{\citenamefont {Croke}\ \emph {et~al.}(2006)\citenamefont {Croke},
  \citenamefont {Andersson}, \citenamefont {Barnett}, \citenamefont {Gilson},\
  and\ \citenamefont {Jeffers}}]{croke2006maximum}%
  \BibitemOpen
  \bibfield  {author} {\bibinfo {author} {\bibfnamefont {S.}~\bibnamefont
  {Croke}}, \bibinfo {author} {\bibfnamefont {E.}~\bibnamefont {Andersson}},
  \bibinfo {author} {\bibfnamefont {S.~M.}\ \bibnamefont {Barnett}}, \bibinfo
  {author} {\bibfnamefont {C.~R.}\ \bibnamefont {Gilson}},\ and\ \bibinfo
  {author} {\bibfnamefont {J.}~\bibnamefont {Jeffers}},\ }\bibfield  {title}
  {\bibinfo {title} {Maximum confidence quantum measurements},\ }\href@noop {}
  {\bibfield  {journal} {\bibinfo  {journal} {Phys. Rev. Lett.}\ }\textbf
  {\bibinfo {volume} {96}},\ \bibinfo {pages} {070401} (\bibinfo {year}
  {2006})}\BibitemShut {NoStop}%
\bibitem [{\citenamefont {Bagan}\ \emph {et~al.}(2012)\citenamefont {Bagan},
  \citenamefont {{Mu\~noz}-Tapia}, \citenamefont {Olivares-Renter\'{\i}a},\
  and\ \citenamefont {Bergou}}]{bagan2012Qrate}%
  \BibitemOpen
  \bibfield  {author} {\bibinfo {author} {\bibfnamefont {E.}~\bibnamefont
  {Bagan}}, \bibinfo {author} {\bibfnamefont {R.}~\bibnamefont
  {{Mu\~noz}-Tapia}}, \bibinfo {author} {\bibfnamefont {G.~A.}\ \bibnamefont
  {Olivares-Renter\'{\i}a}},\ and\ \bibinfo {author} {\bibfnamefont {J.~A.}\
  \bibnamefont {Bergou}},\ }\bibfield  {title} {\bibinfo {title} {Optimal
  discrimination of quantum states with a fixed rate of inconclusive
  outcomes},\ }\href {https://doi.org/10.1103/PhysRevA.86.040303} {\bibfield
  {journal} {\bibinfo  {journal} {Phys. Rev. A}\ }\textbf {\bibinfo {volume}
  {86}},\ \bibinfo {pages} {040303} (\bibinfo {year} {2012})}\BibitemShut
  {NoStop}%
\bibitem [{\citenamefont {Singal}\ \emph {et~al.}(2019)\citenamefont {Singal},
  \citenamefont {Kim},\ and\ \citenamefont {Ghosh}}]{singal2019MEindependent}%
  \BibitemOpen
  \bibfield  {author} {\bibinfo {author} {\bibfnamefont {T.}~\bibnamefont
  {Singal}}, \bibinfo {author} {\bibfnamefont {E.}~\bibnamefont {Kim}},\ and\
  \bibinfo {author} {\bibfnamefont {S.}~\bibnamefont {Ghosh}},\ }\bibfield
  {title} {\bibinfo {title} {Structure of minimum error discrimination for
  linearly independent states},\ }\href
  {https://doi.org/10.1103/PhysRevA.99.052334} {\bibfield  {journal} {\bibinfo
  {journal} {Phys. Rev. A}\ }\textbf {\bibinfo {volume} {99}},\ \bibinfo
  {pages} {052334} (\bibinfo {year} {2019})}\BibitemShut {NoStop}%
\bibitem [{\citenamefont {Ban}\ \emph {et~al.}(1997)\citenamefont {Ban},
  \citenamefont {Kurokawa}, \citenamefont {Momose},\ and\ \citenamefont
  {Hirota}}]{ban1997optimum}%
  \BibitemOpen
  \bibfield  {author} {\bibinfo {author} {\bibfnamefont {M.}~\bibnamefont
  {Ban}}, \bibinfo {author} {\bibfnamefont {K.}~\bibnamefont {Kurokawa}},
  \bibinfo {author} {\bibfnamefont {R.}~\bibnamefont {Momose}},\ and\ \bibinfo
  {author} {\bibfnamefont {O.}~\bibnamefont {Hirota}},\ }\bibfield  {title}
  {\bibinfo {title} {Optimum measurements for discrimination among symmetric
  quantum states and parameter estimation},\ }\href@noop {} {\bibfield
  {journal} {\bibinfo  {journal} {International Journal of Theoretical
  Physics}\ }\textbf {\bibinfo {volume} {36}},\ \bibinfo {pages} {1269}
  (\bibinfo {year} {1997})}\BibitemShut {NoStop}%
\bibitem [{\citenamefont {Barnett}(2001)}]{barnett2001symmetric}%
  \BibitemOpen
  \bibfield  {author} {\bibinfo {author} {\bibfnamefont {S.~M.}\ \bibnamefont
  {Barnett}},\ }\bibfield  {title} {\bibinfo {title} {Minimum-error
  discrimination between multiply symmetric states},\ }\href
  {https://doi.org/10.1103/PhysRevA.64.030303} {\bibfield  {journal} {\bibinfo
  {journal} {Phys. Rev. A}\ }\textbf {\bibinfo {volume} {64}},\ \bibinfo
  {pages} {030303} (\bibinfo {year} {2001})}\BibitemShut {NoStop}%
\bibitem [{\citenamefont {Krovi}\ \emph {et~al.}(2015)\citenamefont {Krovi},
  \citenamefont {Guha}, \citenamefont {Dutton},\ and\ \citenamefont
  {da~Silva}}]{krovi2015symmetric}%
  \BibitemOpen
  \bibfield  {author} {\bibinfo {author} {\bibfnamefont {H.}~\bibnamefont
  {Krovi}}, \bibinfo {author} {\bibfnamefont {S.}~\bibnamefont {Guha}},
  \bibinfo {author} {\bibfnamefont {Z.}~\bibnamefont {Dutton}},\ and\ \bibinfo
  {author} {\bibfnamefont {M.~P.}\ \bibnamefont {da~Silva}},\ }\bibfield
  {title} {\bibinfo {title} {Optimal measurements for symmetric quantum states
  with applications to optical communication},\ }\href
  {https://doi.org/10.1103/PhysRevA.92.062333} {\bibfield  {journal} {\bibinfo
  {journal} {Phys. Rev. A}\ }\textbf {\bibinfo {volume} {92}},\ \bibinfo
  {pages} {062333} (\bibinfo {year} {2015})}\BibitemShut {NoStop}%
\bibitem [{\citenamefont {Barnett}\ and\ \citenamefont
  {Croke}(2009)}]{Barnett09QD}%
  \BibitemOpen
  \bibfield  {author} {\bibinfo {author} {\bibfnamefont {S.~M.}\ \bibnamefont
  {Barnett}}\ and\ \bibinfo {author} {\bibfnamefont {S.}~\bibnamefont
  {Croke}},\ }\bibfield  {title} {\bibinfo {title} {Quantum state
  discrimination},\ }\href@noop {} {\bibfield  {journal} {\bibinfo  {journal}
  {Adv. Opt. Photon.}\ }\textbf {\bibinfo {volume} {1}},\ \bibinfo {pages}
  {238} (\bibinfo {year} {2009})}\BibitemShut {NoStop}%
\bibitem [{\citenamefont {Chefles}(2000)}]{Chefles2000QDReview}%
  \BibitemOpen
  \bibfield  {author} {\bibinfo {author} {\bibfnamefont {A.}~\bibnamefont
  {Chefles}},\ }\bibfield  {title} {\bibinfo {title} {Quantum state
  discrimination},\ }\href {https://doi.org/10.1080/00107510010002599}
  {\bibfield  {journal} {\bibinfo  {journal} {Contemporary Physics}\ }\textbf
  {\bibinfo {volume} {41}},\ \bibinfo {pages} {401} (\bibinfo {year} {2000})},\
  \Eprint {https://arxiv.org/abs/https://doi.org/10.1080/00107510010002599}
  {https://doi.org/10.1080/00107510010002599} \BibitemShut {NoStop}%
\bibitem [{\citenamefont {Bae}\ and\ \citenamefont {Kwek}(2015)}]{Bae_2015}%
  \BibitemOpen
  \bibfield  {author} {\bibinfo {author} {\bibfnamefont {J.}~\bibnamefont
  {Bae}}\ and\ \bibinfo {author} {\bibfnamefont {L.-C.}\ \bibnamefont {Kwek}},\
  }\bibfield  {title} {\bibinfo {title} {Quantum state discrimination and its
  applications},\ }\href {https://doi.org/10.1088/1751-8113/48/8/083001}
  {\bibfield  {journal} {\bibinfo  {journal} {Journal of Physics A:
  Mathematical and Theoretical}\ }\textbf {\bibinfo {volume} {48}},\ \bibinfo
  {pages} {083001} (\bibinfo {year} {2015})}\BibitemShut {NoStop}%
\bibitem [{\citenamefont {Sent\'{\i}s}\ \emph {et~al.}(2016)\citenamefont
  {Sent\'{\i}s}, \citenamefont {Bagan}, \citenamefont {Calsamiglia},
  \citenamefont {Chiribella},\ and\ \citenamefont
  {{Mu\~noz}-Tapia}}]{Sentis2016QCP}%
  \BibitemOpen
  \bibfield  {author} {\bibinfo {author} {\bibfnamefont {G.}~\bibnamefont
  {Sent\'{\i}s}}, \bibinfo {author} {\bibfnamefont {E.}~\bibnamefont {Bagan}},
  \bibinfo {author} {\bibfnamefont {J.}~\bibnamefont {Calsamiglia}}, \bibinfo
  {author} {\bibfnamefont {G.}~\bibnamefont {Chiribella}},\ and\ \bibinfo
  {author} {\bibfnamefont {R.}~\bibnamefont {{Mu\~noz}-Tapia}},\ }\bibfield
  {title} {\bibinfo {title} {Quantum change point},\ }\href@noop {} {\bibfield
  {journal} {\bibinfo  {journal} {Phys. Rev. Lett.}\ }\textbf {\bibinfo
  {volume} {117}},\ \bibinfo {pages} {150502} (\bibinfo {year}
  {2016})}\BibitemShut {NoStop}%
\bibitem [{\citenamefont {Sent\'{\i}s}\ \emph {et~al.}(2017)\citenamefont
  {Sent\'{\i}s}, \citenamefont {Calsamiglia},\ and\ \citenamefont
  {{Mu\~noz}-Tapia}}]{QCPUnAmb}%
  \BibitemOpen
  \bibfield  {author} {\bibinfo {author} {\bibfnamefont {G.}~\bibnamefont
  {Sent\'{\i}s}}, \bibinfo {author} {\bibfnamefont {J.}~\bibnamefont
  {Calsamiglia}},\ and\ \bibinfo {author} {\bibfnamefont {R.}~\bibnamefont
  {{Mu\~noz}-Tapia}},\ }\bibfield  {title} {\bibinfo {title} {Exact
  identification of a quantum change point},\ }\href@noop {} {\bibfield
  {journal} {\bibinfo  {journal} {Phys. Rev. Lett.}\ }\textbf {\bibinfo
  {volume} {119}},\ \bibinfo {pages} {140506} (\bibinfo {year}
  {2017})}\BibitemShut {NoStop}%
\bibitem [{\citenamefont {Sent\'{\i}s}\ \emph {et~al.}(2018)\citenamefont
  {Sent\'{\i}s}, \citenamefont {Mart\'{\i}nez-Vargas},\ and\ \citenamefont
  {{Mu\~noz}-Tapia}}]{Sentis2019Online}%
  \BibitemOpen
  \bibfield  {author} {\bibinfo {author} {\bibfnamefont {G.}~\bibnamefont
  {Sent\'{\i}s}}, \bibinfo {author} {\bibfnamefont {E.}~\bibnamefont
  {Mart\'{\i}nez-Vargas}},\ and\ \bibinfo {author} {\bibfnamefont
  {R.}~\bibnamefont {{Mu\~noz}-Tapia}},\ }\bibfield  {title} {\bibinfo {title}
  {Online strategies for exactly identifying a quantum change point},\
  }\href@noop {} {\bibfield  {journal} {\bibinfo  {journal} {Phys. Rev. A}\
  }\textbf {\bibinfo {volume} {98}},\ \bibinfo {pages} {052305} (\bibinfo
  {year} {2018})}\BibitemShut {NoStop}%
\bibitem [{\citenamefont {Skotiniotis}\ \emph {et~al.}(2018)\citenamefont
  {Skotiniotis}, \citenamefont {{Hotz}}, \citenamefont {{Calsamiglia}},\ and\
  \citenamefont {{Mu{\~n}oz-Tapia}}}]{Identification2018Skotiniotis}%
  \BibitemOpen
  \bibfield  {author} {\bibinfo {author} {\bibfnamefont {M.}~\bibnamefont
  {Skotiniotis}}, \bibinfo {author} {\bibfnamefont {R.}~\bibnamefont {{Hotz}}},
  \bibinfo {author} {\bibfnamefont {J.}~\bibnamefont {{Calsamiglia}}},\ and\
  \bibinfo {author} {\bibfnamefont {R.}~\bibnamefont {{Mu{\~n}oz-Tapia}}},\
  }\bibfield  {title} {\bibinfo {title} {Identification of malfunctioning
  quantum devices},\ }\href@noop {} {\bibfield  {journal} {\bibinfo  {journal}
  {arXiv:1808.02729}\ } (\bibinfo {year} {2018})}\BibitemShut {NoStop}%
\bibitem [{\citenamefont {Nielsen}\ and\ \citenamefont
  {Chuang}(2011)}]{Nielsen2011QCQ}%
  \BibitemOpen
  \bibfield  {author} {\bibinfo {author} {\bibfnamefont {M.~A.}\ \bibnamefont
  {Nielsen}}\ and\ \bibinfo {author} {\bibfnamefont {I.~L.}\ \bibnamefont
  {Chuang}},\ }\href@noop {} {\emph {\bibinfo {title} {Quantum Computation and
  Quantum Information: 10th Anniversary Edition}}},\ \bibinfo {edition} {10th}\
  ed.\ (\bibinfo  {publisher} {Cambridge University Press},\ \bibinfo {address}
  {New York, NY, USA},\ \bibinfo {year} {2011})\BibitemShut {NoStop}%
\bibitem [{\citenamefont {Vandenberghe}\ and\ \citenamefont
  {Boyd}(1996)}]{VandenbergheSDP1996}%
  \BibitemOpen
  \bibfield  {author} {\bibinfo {author} {\bibfnamefont {L.}~\bibnamefont
  {Vandenberghe}}\ and\ \bibinfo {author} {\bibfnamefont {S.}~\bibnamefont
  {Boyd}},\ }\bibfield  {title} {\bibinfo {title} {Semidefinite programming},\
  }\href@noop {} {\bibfield  {journal} {\bibinfo  {journal} {SIAM Review}\
  }\textbf {\bibinfo {volume} {38}},\ \bibinfo {pages} {49} (\bibinfo {year}
  {1996})}\BibitemShut {NoStop}%
\bibitem [{\citenamefont {{Eldar}}(2003)}]{Eldar2002ASDP}%
  \BibitemOpen
  \bibfield  {author} {\bibinfo {author} {\bibfnamefont {Y.~C.}\ \bibnamefont
  {{Eldar}}},\ }\bibfield  {title} {\bibinfo {title} {A semidefinite
  programming approach to optimal unambiguous discrimination of quantum
  states},\ }\href@noop {} {\bibfield  {journal} {\bibinfo  {journal} {IEEE
  Transactions on Information Theory}\ }\textbf {\bibinfo {volume} {49}},\
  \bibinfo {pages} {446} (\bibinfo {year} {2003})}\BibitemShut {NoStop}%
\bibitem [{\citenamefont {Watrous}(2018)}]{watrous_2018}%
  \BibitemOpen
  \bibfield  {author} {\bibinfo {author} {\bibfnamefont {J.}~\bibnamefont
  {Watrous}},\ }\href@noop {} {\emph {\bibinfo {title} {The Theory of Quantum
  Information}}}\ (\bibinfo  {publisher} {Cambridge University Press},\
  \bibinfo {year} {2018})\BibitemShut {NoStop}%
\bibitem [{\citenamefont {Bhatia}(1996)}]{bhatia1996matrix}%
  \BibitemOpen
  \bibfield  {author} {\bibinfo {author} {\bibfnamefont {R.}~\bibnamefont
  {Bhatia}},\ }\href@noop {} {\emph {\bibinfo {title} {Matrix Analysis}}},\
  Graduate Texts in Mathematics\ (\bibinfo  {publisher} {Springer New York},\
  \bibinfo {year} {1996})\BibitemShut {NoStop}%
\bibitem [{\citenamefont {Horn}\ and\ \citenamefont
  {Johnson}(2012)}]{HornMA2012}%
  \BibitemOpen
  \bibfield  {author} {\bibinfo {author} {\bibfnamefont {R.~A.}\ \bibnamefont
  {Horn}}\ and\ \bibinfo {author} {\bibfnamefont {C.~R.}\ \bibnamefont
  {Johnson}},\ }\href@noop {} {\emph {\bibinfo {title} {Matrix Analysis}}},\
  \bibinfo {edition} {2nd}\ ed.\ (\bibinfo  {publisher} {Cambridge University
  Press},\ \bibinfo {address} {New York, NY, USA},\ \bibinfo {year}
  {2012})\BibitemShut {NoStop}%
\bibitem [{\citenamefont {Hausladen}\ and\ \citenamefont
  {Wootters}(1994)}]{Hausladen94PrettyGood}%
  \BibitemOpen
  \bibfield  {author} {\bibinfo {author} {\bibfnamefont {P.}~\bibnamefont
  {Hausladen}}\ and\ \bibinfo {author} {\bibfnamefont {W.~K.}\ \bibnamefont
  {Wootters}},\ }\bibfield  {title} {\bibinfo {title} {A `pretty good'
  measurement for distinguishing quantum states},\ }\href@noop {} {\bibfield
  {journal} {\bibinfo  {journal} {Journal of Modern Optics}\ }\textbf {\bibinfo
  {volume} {41}},\ \bibinfo {pages} {2385} (\bibinfo {year}
  {1994})}\BibitemShut {NoStop}%
\bibitem [{\citenamefont {Hausladen}\ \emph {et~al.}(1996)\citenamefont
  {Hausladen}, \citenamefont {Jozsa}, \citenamefont {Schumacher}, \citenamefont
  {Westmoreland},\ and\ \citenamefont {Wootters}}]{Hausladen96Classical}%
  \BibitemOpen
  \bibfield  {author} {\bibinfo {author} {\bibfnamefont {P.}~\bibnamefont
  {Hausladen}}, \bibinfo {author} {\bibfnamefont {R.}~\bibnamefont {Jozsa}},
  \bibinfo {author} {\bibfnamefont {B.}~\bibnamefont {Schumacher}}, \bibinfo
  {author} {\bibfnamefont {M.}~\bibnamefont {Westmoreland}},\ and\ \bibinfo
  {author} {\bibfnamefont {W.~K.}\ \bibnamefont {Wootters}},\ }\bibfield
  {title} {\bibinfo {title} {Classical information capacity of a quantum
  channel},\ }\href@noop {} {\bibfield  {journal} {\bibinfo  {journal} {Phys.
  Rev. A}\ }\textbf {\bibinfo {volume} {54}},\ \bibinfo {pages} {1869}
  (\bibinfo {year} {1996})}\BibitemShut {NoStop}%
\bibitem [{\citenamefont {Dalla~Pozza}\ and\ \citenamefont
  {Pierobon}(2015)}]{DallaPozza2015Optimality}%
  \BibitemOpen
  \bibfield  {author} {\bibinfo {author} {\bibfnamefont {N.}~\bibnamefont
  {Dalla~Pozza}}\ and\ \bibinfo {author} {\bibfnamefont {G.}~\bibnamefont
  {Pierobon}},\ }\bibfield  {title} {\bibinfo {title} {Optimality of
  square-root measurements in quantum state discrimination},\ }\href@noop {}
  {\bibfield  {journal} {\bibinfo  {journal} {Phys. Rev. A}\ }\textbf {\bibinfo
  {volume} {91}},\ \bibinfo {pages} {042334} (\bibinfo {year}
  {2015})}\BibitemShut {NoStop}%
\bibitem [{\citenamefont {Gray}(2006)}]{Gray2006Toeplitz}%
  \BibitemOpen
  \bibfield  {author} {\bibinfo {author} {\bibfnamefont {R.~M.}\ \bibnamefont
  {Gray}},\ }\bibfield  {title} {\bibinfo {title} {Toeplitz and circulant
  matrices: A review},\ }\href@noop {} {\bibfield  {journal} {\bibinfo
  {journal} {Foundations and Trends in Communications and Information Theory}\
  }\textbf {\bibinfo {volume} {2}},\ \bibinfo {pages} {155} (\bibinfo {year}
  {2006})}\BibitemShut {NoStop}%
\bibitem [{\citenamefont {Katznelson}(2004)}]{Katznelson2002}%
  \BibitemOpen
  \bibfield  {author} {\bibinfo {author} {\bibfnamefont {Y.}~\bibnamefont
  {Katznelson}},\ }\href@noop {} {\emph {\bibinfo {title} {An Introduction To
  Harmonic Analysis}}},\ Cambridge Mathematical Library\ (\bibinfo  {publisher}
  {Cambridge University Press},\ \bibinfo {year} {2004})\BibitemShut {NoStop}%
\end{thebibliography}%
\end{document}